# Z Cam Stars in the Twenty-First Century

**Mike Simonsen**
*AAVSO, 49 Bay State Road., Cambridge, MA 02138; mikesimonsen@aavso.org*

**David Boyd**
*5 Silver Lane, West Challow OX12 9TX, England; davidboyd@orion.me.uk*

**William Goff**
*13508 Monitor Lane, Sutter Creek, CA 95685; b-goff@sbcglobal.net*

**Tom Krajci**
*Center for Backyard Astronomy, P.O. Box 1351, Cloudcroft, NM 88317; tom_krajci@tularosa.net*

**Kenneth Menzies**
*318A Potter Road, Framingham MA, 01701; kenmenstar@gmail.com*

**Sebastian Otero**
*AAVSO, 49 Bay State Road, Cambridge, MA 02138; sebastian@aavso.org*

**Stefano Padovan**
*Barrio Masos SN, 17132 Foixà, Girona, Spain; stefano@stefanopadovan.com*

**Gary Poyner**
*67 Ellerton Road, Kingstanding, Birmingham B44 0QE, England; garypoyner@blueyonder.co.uk*

**James Roe**
*85 Eikermann Road-174, Bourbon, MO 65441; jim.roe@asemonline.org*

**Richard Sabo**
*2336 Trailcrest Drive, Bozeman, MT 59718; rsabo333@gmail.com*

**George Sjoberg**
*9 Contentment Crest, #182, Mayhill, NM 88339; george.y.sjoberg@gmail.com*

**Bart Staels**
*Koningshofbaan 51, Hofstade (Aalst) B-9308, Belgium; staels.bart.bvba@pandora.be*

**Rod Stubbings**
*2643 Warragul, Korumburra Road, Tetoora Road, VIC 3821, Australia; stubbo@sympac.com.au*

**John Toone**
*17 Ashdale Road, Cressage, Shrewsbury SY5 6DT, England; enootnhoj@btinternet.com*

**Patrick Wils**
*Aarschotsebaan 31, Hever B-3191, Belgium; patrickwils@yahoo.com*

**Abstract**  Z Cam (UGZ) stars are a small subset of dwarf novae that exhibit standstills in their light curves. Most modern literature and catalogs of cataclysmic variables quote the number of known Z Cams to be on the order of thirty or so systems. After a four-year observing campaign and an exhaustive examination of the data in the AAVSO International Database (AID) we have trimmed that number by a third. One of the reasons for the misclassification of some systems is the fact that the definition of what a Z Cam is has changed over the last eighty-five years. This has caused many stars formerly assumed to be Z Cams or rumored to be Z Cams to be eliminated from the final list. In this paper we present the results of our investigation into sixty-five stars listed at one time or another in the literature as Z Cams or possible Z Cams.

## 1. Introduction

Dwarf novae (DNe), or U Geminorum (UG) systems, are a subclass of cataclysmic variable stars (CVs), semi-detached close binary systems in which a white dwarf (WD) accretes material from a Roche-lobe-filling secondary via an accretion disk (Warner 1995). They are called "dwarf novae" because DNe outbursts are smaller in amplitude and higher in frequency than classical novae eruptions.

DNe outbursts result from a temporary increase in the rate of accretion onto the WD. According to the thermal-viscous disc instability model, over time the mass of accreted material grows and the temperature of the disk rises, until it becomes sufficiently high to switch into a hot, highly viscous state. The disk becomes unstable and mass is dumped onto the surface of the WD after plowing through a violent transition region just above the surface of the WD called the "boundary layer," releasing copious amounts of energy from optical to X-rays.

DNe are classified into sub-groups based primarily on their light curves. SS Cygni stars (UGSS) brighten dramatically by 2 to 6 magnitudes in 1 to 2 days, and return to their original brightness after a period of several days to a week or more. Cycle times between outbursts range from 10 days to years. Orbital periods are usually longer than 3 hours. SU Ursae Majoris stars (UGSU) exhibit normal outbursts, but also have brighter and longer "superoutbursts." The cycle times of superoutbursts (super-cycle) are usually several times the length of time between normal outbursts and can be years or decades long. In general, the shorter the orbital period, the longer the super-cycle. Orbital periods are usually in the range of 75 minutes to 2 hours. Z Cam stars (UGZ) are DNe that exhibit normal UG-type outbursts, as well as random standstills. A standstill usually starts at the end of an outburst and consists of a period of relatively constant brightness 1 to 1.5 magnitudes below maximum light that may last from a few days to 1,000 days. The orbital periods of UGZ are all longer than 3 hours.

This classification system has evolved over time as our understanding of the physical processes behind the observed behaviors has grown. However, the changing definition of UGZ stars has created confusion in the literature and many stars that used to fit the definition no longer meet the requirements of membership in the Z Cam classification.

**2. Origin and evolution of the Z Cam classification**

The English astronomer J. R. Hind discovered U Gem on December 15, 1855. The speed and amplitude of its brightness variations attracted the attention of many of the leading observers of the day, and for over forty years it was the only star of its type known. Then, in 1896, Miss Louisa D. Wells discovered SS Cyg on plates taken at the Harvard College Observatory. The similarity between these two stars was noted immediately and the classification of U Gem stars was introduced. The early definition was based on light curves of stars that stayed at minimum for the majority of the time, but at intervals between 40 and 100 days erupted by 3 to 5 magnitudes (Toone 2012). By 1928 there were two dozen or more stars classified as UG or suspected of being UG.

Two stars, Z Cam and RX And, discovered in 1904 and 1905, respectively, were initially classified as UG, but years of close observation revealed these two stars to have much shorter periods (at that time, defined as the time between maxima) and they spent very little time at minimum. This caused A. A. Nijland to propose a new class of variable stars, the Z Cam type, of which he suggested X Leo and TW Vir might also be members (Nijland 1930). This new classification was further bolstered by support from Felix De Roy, Director of the British Astronomical Association-Variable Star Section (BAA-VSS), when he published a paper, "A New Variable Star Class, The Z Camelopardalis Type," describing the state of knowledge of these stars to date (De Roy 1932).

The roots of some of the confusion on the classification of several stars can be traced back to this paper in which RX And, Z Cam, BI Ori, CN Ori, TZ Per, SV CMi, X Leo, and SU UMa are either included in the new class or suspected of membership and requiring further investigation.

It is also this paper that first defined the "crucial features for the Z Cam type." In 1932 they were:

1. The short duration of minimum.

2. The irregularity of the light curve, described as rare for U Gem types and almost the norm for Z Cams.

3. The lesser amplitudes of variation compared to U Gems, 2.64 magnitudes for Z Cams versus 3.8 magnitudes for U Gems.

4. A "curious and very special feature" where the variable remains nearly constant at a magnitude in between the maximum and minimum.

Interestingly, it is mentioned that only Z Cam, TZ Per, and RX And exhibit this feature. At the time these "standstills," as they would come to be known, were not the primary feature of Z Cam stars, but more of a curiosity. It was the brief minima and short duration between maxima that set these stars apart initially.

Campbell and Jacchia (1941) note in *The Story of Variable Stars*, "from time to time they take a sort of vacation, and remain at almost constant brightness." But this is a footnote to the description based primarily on the hyperactive nature of Z Cams.

The standstills take on more prominence in Elvey and Babcock (1943) where they write, "Whenever they go through their regular variations, they behave similarly to the short period group of SS Cygni stars. However, these stars may remain for weeks at relatively constant brightness, approximately one-third from maximum to minimum brightness."

By 1971, the term "standstills" was in use and is described in *The Variable Star Observers Handbook* (Glasby 1971) as the main distinguishing feature of Z Cam type variables. "The major difference, and that which justifies their inclusion in a separate group, is the periods of standstill."

The modern day definition in the *General Catalogue of Variable Stars* (GCVS; Samus *et al.* 2007–2009) also stresses the importance of standstills as the determining characteristic of Z Cams:

> Z Camelopardalis type stars. These also show cyclic outbursts, differing from UGSS variables by the fact that sometimes after an outburst they do not return to the original brightness, but during several cycles retain a magnitude between maximum and minimum. The values of cycles are from 10 to 40 days, while light amplitudes are from 2 to 5 magnitudes in V.

This was the definition used when the Z CamPaign was launched in 2009 (Simonsen 2011), in order to differentiate between genuine Z Cam stars and their imposters. The criterion for inclusion in the Z Cam class was simply evidence of at least one standstill in the available data.

For many potential Z Cam stars, significant data exist in the AAVSO International Database which can be used to determine whether they belong to this class or not. For many others, the challenge has been to acquire enough data over a multi-year campaign to weed out the pretenders from the bona fide Z Cams.

## 3. Bona fide Z Cam stars

After four years of Z CamPaign observations and an exhaustive examination of the AAVSO International Database, nineteen bona fide Z Cam type systems have been identified out of the sixty-five stars listed in the literature at one time or another as Z Cam stars. Some of these have been known to be Z Cams for decades. Others are newly classified or re-classified as Z Cams.

All light curves (Figures 1–21) in this section were created using vstar (Benn 2012). Black dots are AAVSO visual data; green dots [gray in black/white version] are AAVSO Johnson V data (AAVSO 2013). Except for the prototype star Z Cam, the stars are listed here in order of R.A. Each star's position, maximum and minimum, and type is given in Table 1.

The entire set of AAVSO data for each star was downloaded and the light curve was examined visually, both in gross, multi-year displays and in close detail, cycle-by-cycle. Start dates of the data set and the beginnings and endings of outbursts and standstills could be determined by selecting individual data points on the vstar light curve and viewing the complete observational information on the vstar information screen.

Table 1: Bona fide Z Cam–type variable stars.

| Name | R.A. (2000) h m s | Dec. (2000) ° ′ ″ | Max | Min | Type |
|---|---|---|---|---|---|
| WW Cet | 00 11 24.78 | −11 28 43.1 | 10.4V | 15.8V | UGZ |
| V513 Cas | 00 18 14.91 | 66 18 13.6 | 14.9V | 18.9V | UGZ* |
| IW And | 01 01 08.91 | 43 23 25.7 | 13.8V | 17.7V | UGZ* |
| RX And | 01 04 35.54 | 41 17 57.8 | 10.2V | 15.1V | UGZ |
| AY Psc | 01 36 55.45 | 07 16 29.3 | 15.3V | 17.0V | UGZ+E |
| TZ Per | 02 13 50.97 | 58 22 52.3 | 12.0V | 15.6V | UGZ |
| HL CMa | 06 45 17.22 | −16 51 34.7 | 10.6V | 14.9V | UGZ* |
| BX Pup | 07 54 15.57 | −24 19 36.5 | 13.8V | 18.0V | UGZ |
| Z Cam | 08 25 13.20 | 73 06 39.2 | 10.0V | 14.5V | UGZ* |
| AT Cnc | 08 28 36.93 | 25 20 03.0 | 12.7V | 15.2V | UGZ* |
| SY Cnc | 09 01 03.32 | 17 53 56.2 | 11.1V | 14.5V | UGZ |
| Leo5 | 10 28 00.08 | 21 48 13.7 | 15.2V | 17.7V | UGZ+E* |
| ES Dra | 15 25 31.81 | 62 01 00.0 | 14.1V | 17.5V | UGZ |
| HP Nor | 16 20 49.59 | −54 53 22.9 | 12.8v | 16.4V | UGZ |
| AH Her | 16 44 10.01 | 25 15 02.0 | 11.3v | 14.7V | UGZ* |
| UZ Ser | 18 11 24.90 | −14 55 33.9 | 12.4V | 17.2V | UGZ* |
| EM Cyg | 19 38 40.11 | 30 30 28.4 | 12.0V | 14.3V | UGZ+E |
| VW Vul | 20 57 45.07 | 25 30 25.7 | 13.1V | 17.0V | UGZ |
| HX Peg | 23 40 23.70 | 12 37 41.8 | 12.9V | 16.6V | UGZ* |

*Outbursts from standstills have been observed.*

Average cycle times were determined using a custom-designed tool in vstar. Maxima could be hand selected individually, and the mean magnitude and average time between selections were automatically calculated. The averages of the densest data sets were then calculated to derive the mean time between selections and the mean magnitude of maxima.

3.1. Z Cam (the prototype star)

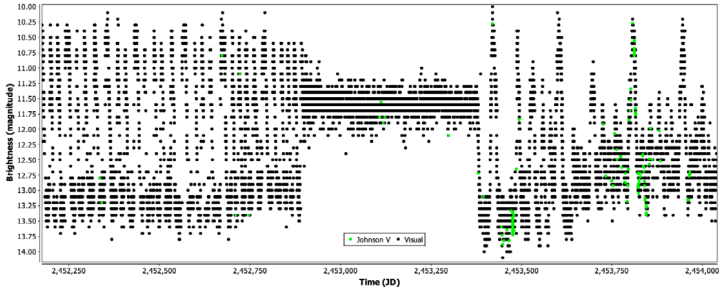

Figure 1. AAVSO light curve for Z Cam, the prototype of the class, showing the normal rapid succession of outburst to quiescence interrupted by a standstill from JD 2452896 to 2453381 (September 2003 to January 2005). There is then a drop to quiescence followed by an irregular pattern of outbursts.

3.2. WW Cet

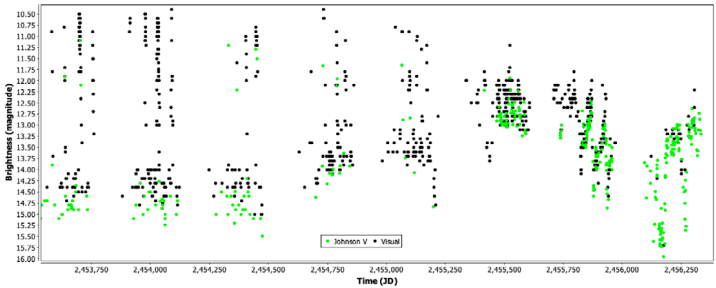

Figure 2. The AAVSO light curve of WW Cet shows a normal pattern of outbursts and minima, followed by the first standstill in AAVSO records, from JD 2455359 to 2455826 (June 2010 to September 2011). This is followed by a weaker pattern of outbursts with progressively fainter minima. Recently classified as Z Cam in Simonsen and Stubbings (2011).

### 3.3. RX And

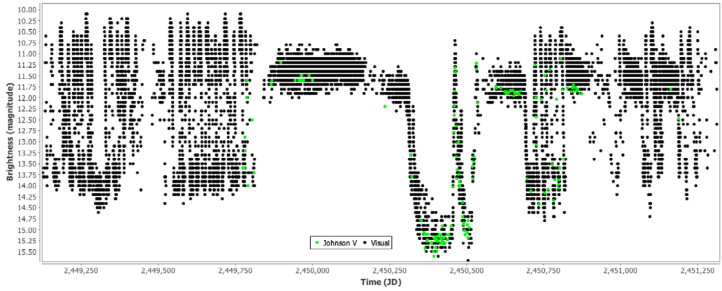

Figure 3. The AAVSO light curve for RX And, one of the first Z Cams classified as such (Nijland 1930; De Roy 1932), clearly shows a standstill from JD 2449843 to 2450315 (May 1995 to August of 1996), followed by a precipitous drop to an uncharacteristically faint minimum. This is followed by two brief outbursts, and then a series of standstills and outbursts.

### 3.4. AY Psc

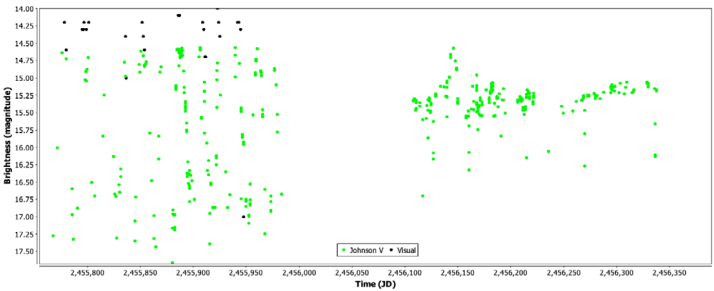

Figure 4. Classified as a Z Cam in Mercado and Honeycutt (2002), this AAVSO light curve for AY Psc shows a normal pattern of maximum and minimum cycles, then a seasonal gap in the data. This is followed by progressively brighter minima and a standstill from JD 2456163 to 2456335 (August 2012 to February 2013).

### 3.5. TZ Per

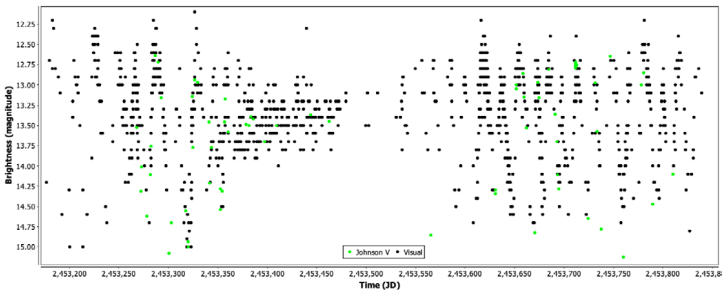

Figure 5. TZ Per is one of the original Z Cams classified in De Roy (1932). This AAVSO light curve shows a regular pattern of outbursts that is interrupted by a standstill from JD 2453394 to 2453511 (January to May 2005). This is followed by a seasonal gap, and then the resumption of normal maximum and minimum cycles.

 Simonsen et al., JAAVSO Volume 42, 2014

### 3.6. HL CMa

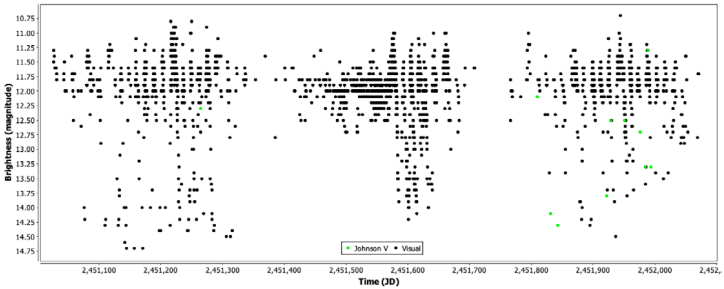

Figure 6. Classified as a Z Cam at least as early as Mansperger *et al.* (1994), this AAVSO light curve for HL CMa shows normal outburst cycles interrupted by a standstill from JD 2451386 to 2451574 (July 1999 to January 2000). This is followed by a rise to maximum before normal maximum and minimum cycles resume.

### 3.7. BX Pup

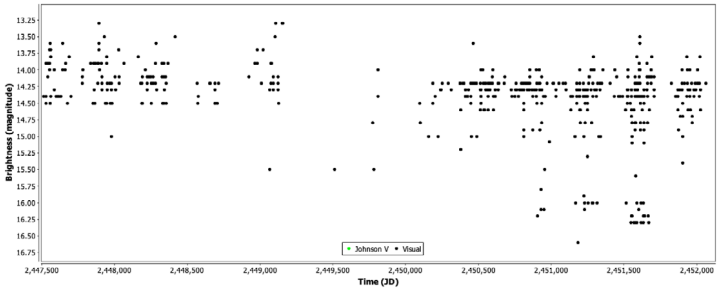

Figure 7. This AAVSO light curve of BX Pup, although sparse, does seem to indicate at least two standstills. The first from JD 2448569 to 2448713 (November 1991 to April 1992) and another more extended standstill from JD 2450097 to 2450904 (January 1996 to March 1998). There is another sparsely sampled standstill from JD 2452339 to 2452747 (April 2002 to April 2003) in the AAVSO data (Figure 8). This, coupled with observations of a 38-day standstill in Dirks (1941), leaves us to conclude that BX Pup is indeed a Z Cam.

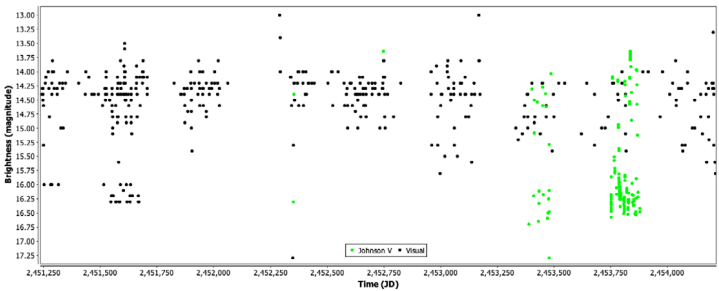

Figure 8. AAVSO light curve of BX Pup centered on the April 2002 to April 2003 standstill (JD 2452339 to 2452747).

### 3.8. AT Cnc

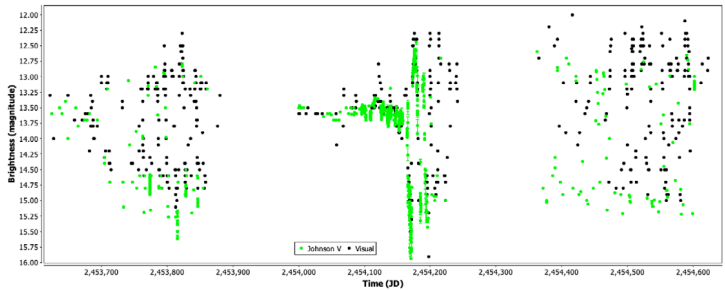

Figure 9. This AAVSO light curve of AT Cnc is a textbook example of a normal outburst cycle, followed by a seasonal gap, then an obvious standstill from JD 2453999 to 2454163 (September 2006 to March 2007), followed by a deep fade to minimum, then the resumption of normal maximum to minimum cycles.

### 3.9. SY Cnc

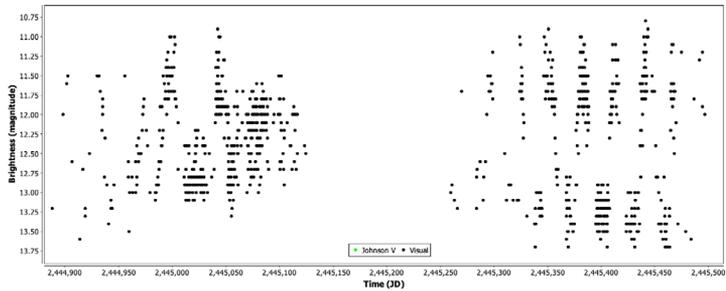

Figure 10. This standstill in SY Cnc, beginning on or around JD 2445069, lasts at least until JD 2445124 (April to June 1982). We don't know how long the standstill continued because the star is lost to the horizon and the seasonal gap. Though truncated by the seasonal gap, this is the only convincing evidence of a standstill in all the AAVSO data. However, there are tantalizing indications that several standstills may have taken place during other seasonal gaps.

### 3.10. Leo5

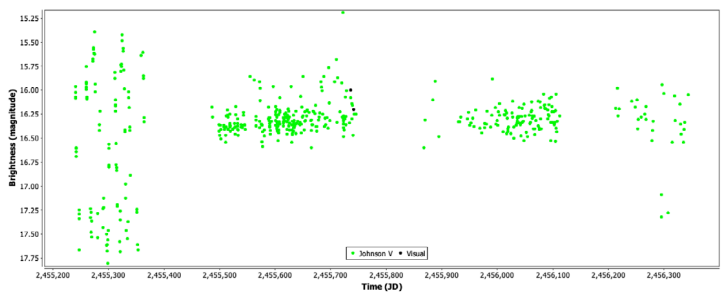

Figure 11. Leo5 is a newly classified Z Cam (Wils *et al.* 2011). AAVSO data for Leo5 show the normal outburst cycle, interrupted by a standstill beginning in October 2010 (JD 2455485), then a brief outburst from standstill in June 2011 (JD 2455721). The standstill then resumes, until it drops to minimum in January 2013 (JD 2456295).

3.11. AH Her

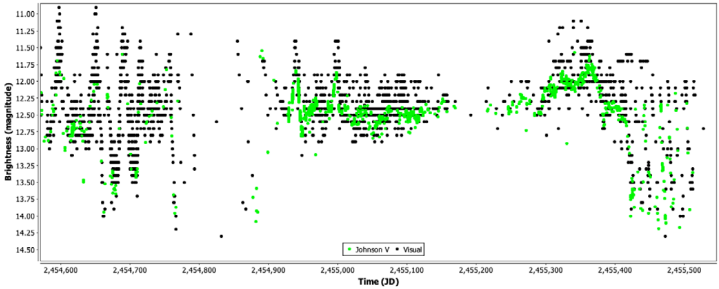

Figure 12. This AAVSO light curve of AH Her shows, in exquisite detail, the star settling into a standstill beginning on JD 2455005 (June 2009). It becomes slightly more chaotic, and then goes into outburst from standstill almost one year later on JD 2455363, before fading to progressively fainter minima.

3.12. UZ Ser

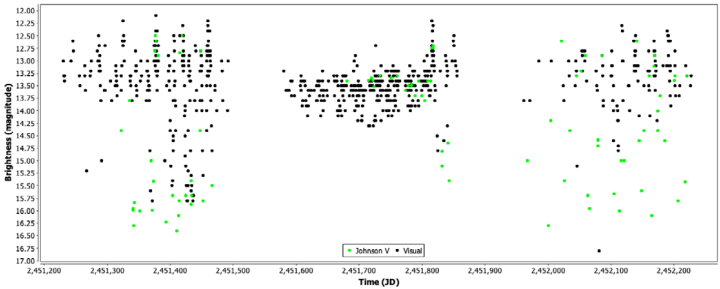

Figure 13. This AAVSO light curve clearly shows UZ Ser in a standstill beginning JD 2451580 (February 2000), followed by an outburst from standstill at JD 2451815 (September 2000). It then resumes more or less normal outburst cycles.

3.13. EM Cyg

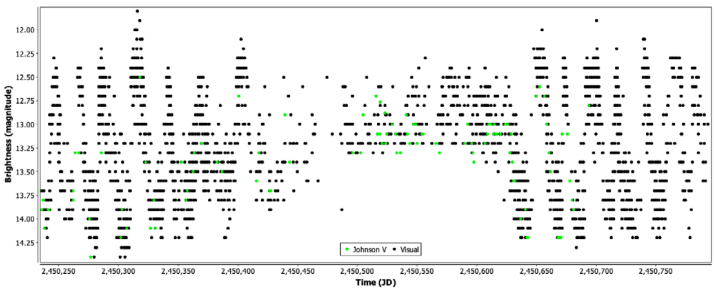

Figure 14. This AAVSO light curve shows EM Cyg in standstill from JD 2450454 to 2450636 (January to July 1997), preceded and followed by more or less normal outburst cycles.

### 3.14. VW Vul

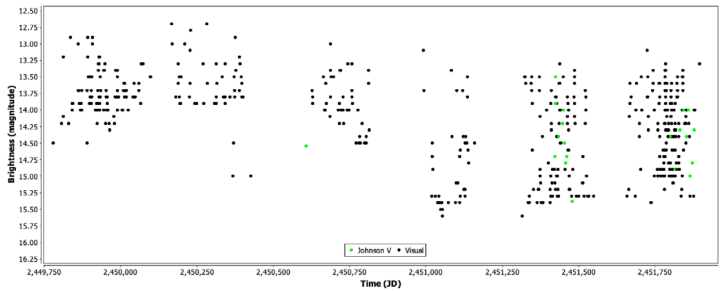

Figure 15. This AAVSO light curve shows VW Vul in standstill from JD 2450627 to 2450814 (June 1997 to January 1998).

### 3.15. HX Peg

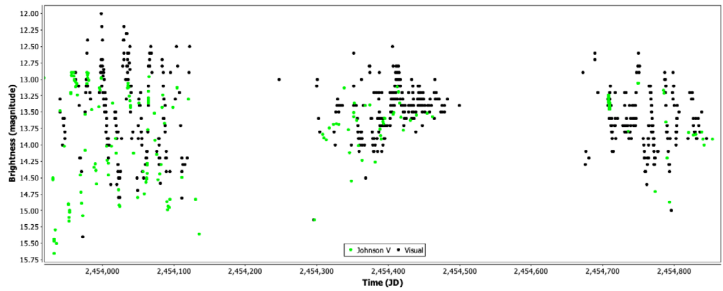

Figure 16. This AAVSO light curve of HX Peg shows it settling down to standstill on or around JD 2454423 (November 2007) until it is lost to the seasonal gap after JD 2454499 (February 2008).

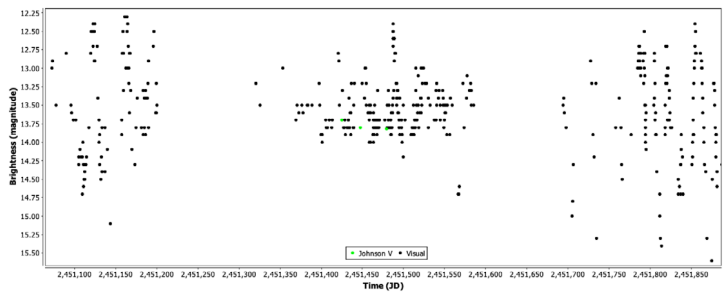

Figure 17. Here we see HX Peg go into outburst in November 1999 (JD 2451487) from standstill, then back into standstill at the end of 1999.

### 3.16. ES Dra

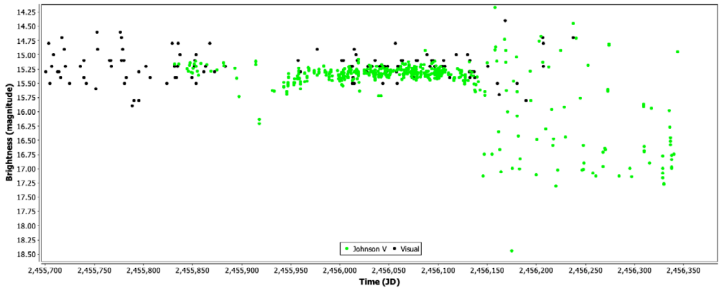

Figure 18. This AAVSO light curve shows the most recent standstill of ES Dra, from JD 2455931 to 2456167 (January to August 2012), firmly establishing ES Dra as a member of the Z Cam class. See also Ringwald and Velasco (2012).

### 3.17. HP Nor

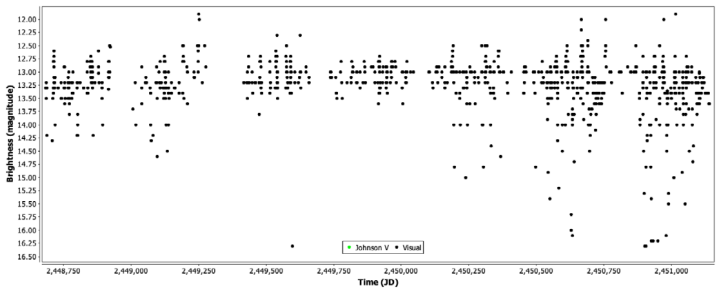

Figure 19. This AAVSO light curve of HP Nor shows a standstill beginning JD 2449737 (January 1995) and lasting until JD 2450161 (March 1996).

### 3.18. IW And

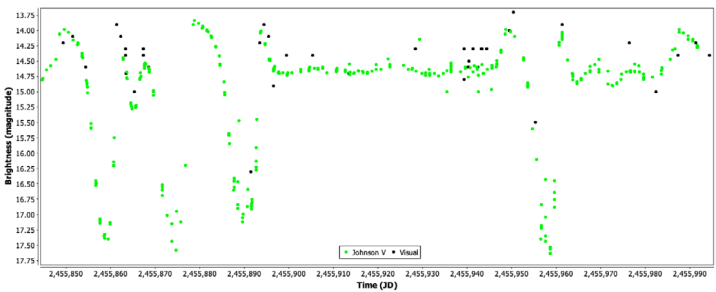

Figure 20. This AAVSO light curve shows IW And in a standstill December 1, 2011 (JD 2456263), followed by a rise to outburst on January 10, 2012 (JD 2456303). This is followed by a fade to minimum, a short outburst, and then another standstill—which is followed by yet another outburst.

### 3.19. V513 Cas

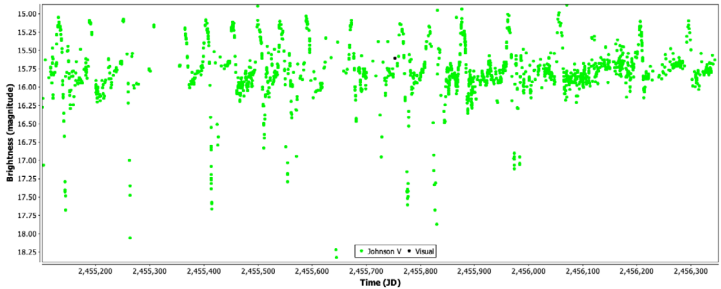

Figure 21. This AAVSO light curve shows the previously unknown, exotic nature of V513 Cas. Beginning in October of 2009 (JD 2455105) the system was observed with a pattern of outbursts every 51 days on average, punctuated by 20- to 30-day standstills, rise to outburst again, and then a deep fading episode roughly every other cycle. In February 2012 the pattern changed dramatically. Cycle time between outbursts stretched to an average of 83 days, the standstills were less chaotic, and there were no deep fades to 17th magnitude, resembling nova-like variable (NL) behavior.

## 4. Characteristics of true Z Cam stars

There are several other characteristics which many Z Cams share in addition to the modern defining characteristic of standstills.

### 4.1. Standstills

As the most significant characteristic of assigning membership to the Z Cam classification of DNe, it is appropriate to begin with our current understanding of Z Cam standstills.

The word "standstill" is somewhat misleading. Their light curves do not look like the flat line of an EKG graph of a patient whose heart has stopped beating. Indeed, Z Cams are quite lively even in standstill. Visual examination of the light curves of standstills reveals a remarkable amount of activity and "jitter." Szkody and Mattei (1984) showed erratic flare-ups with amplitudes up to several tenths of a magnitude in Z Cam standstills.

It is the cause of these episodes of more or less steady light fainter than maximum that has stirred the greatest amount of discussion. It is generally agreed nowadays that standstills are the result of a sudden increase in mass transfer, above the rate allowing for normal SS Cygni-type outburst cycles and below the rate that would cause the system to remain in a state of continuous outburst, like the nova-like stars. What causes this increase has been the subject of much debate for over thirty years.

Based on the disk-instability model, Meyer and Meyer-Hofmeister (1983) proposed that Z Cams normally have mass transfer rates slightly below the critical value that would keep their accretion rates stable. Irradiation of the secondary is given as a reason for the higher mass accretion rate seen in standstills than in outburst cycle phases. They suggest standstills occur when the mass transfer rate changes because of irradiation of the secondary, the mass transfer stream impacting the disk and tidal friction. They also suggest a "relaxation oscillation" cycle that happens as the mass transfer rate drops to lower levels, allowing the outburst cycle to begin again after a standstill.

Oppenheimer *et al.* (1998) argued that irradiation of the secondary does not play a significant role in the changes in mass accretion rates in Z Cams. If this were true, a standstill should accompany a bright quiescence, because an irradiated secondary should be brighter and lose more material into the bright spot. Their analysis showed that faint quiescences accompany standstill intervals. They suggest solar-type magnetic cycles and star spots as a plausible alternative mechanism. Smak (2004) also concludes that irradiation from the secondary is not a significant factor in enhanced mass transfer in Z Cam systems.

Stehle *et al.* (2001) explain standstills are fainter than outburst maxima because the gas stream from the donor star heats the disk, which lowers the threshold of mass transfer needed to keep the star from going back to quiescence. Their model predicts standstill luminosities to be about 40% less than the peak brightness of an outburst, which matches observations very well. Buat-Menard *et al.* (2001) also conclude that better agreement with the observations is obtained when one takes into account the energy released by the impact of the mass transfer stream onto the disk and by tidal torque dissipation.

The one thing none of the models explains is the underlying cause of this sudden shift in the mass transfer rate. What initiates it, and what makes it turn off, allowing the star to go back to quiescence, or in some cases, back into outburst?

An oft-quoted characteristic of Z Cams is that "standstills are always initiated by an outburst," and "standstills always end with a decline to quiescence" (Hellier 2001). However, there are at least nine Z Cam stars that have been shown since since 1959 (Collinson and Isles 1979) to go into outburst from standstill: Z Cam, HX Peg, AH Her, HL CMa, UZ Ser, AT Cnc, Leo5, V513 Cas, and IW And. This inconvenient truth raises even more questions about the cause of standstills. If it is true that the accretion disk has been drained in the plateau phase just before a standstill (as put forth in Oppenheimer *et al.* 1998), then what is the underlying cause of outbursts that occur immediately after standstills?

4.2. Orbital period

17 of the 19 bona fide Z Cams have orbital periods in the literature. All have periods from 3.048 hours (0.127d) to 8.4 hours (0.38d), the average being 5.272 hours (0.2196d). The distribution of orbital periods is shown in Figure 22.

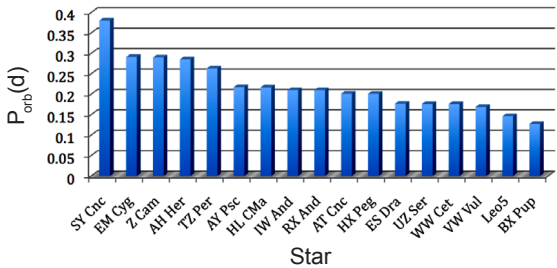

Figure 22. The distribution of orbital periods for 17 of the 19 confirmed Z Cams.

4.3. Outburst cycle

Z Cams are very active systems. Most have outburst cycles (the time between successive maxima) between 10 and 30 days. Their normal cycles between maxima and minima look very much like UGSS stars but they spend very little time at minimum.

4.4. Outburst amplitudes

Outburst amplitudes of Z Cam stars range from 2.3 to 4.9 magnitudes in V (Figure 23). The average amplitude is 3.7V. This is identical to the range of amplitudes seen in UGSS stars, so it cannot be used to distinguish them from these more common DNe. It does set them apart from NLs with smaller-amplitude changes and UGSU and WZ Sge stars with larger-amplitude outbursts.

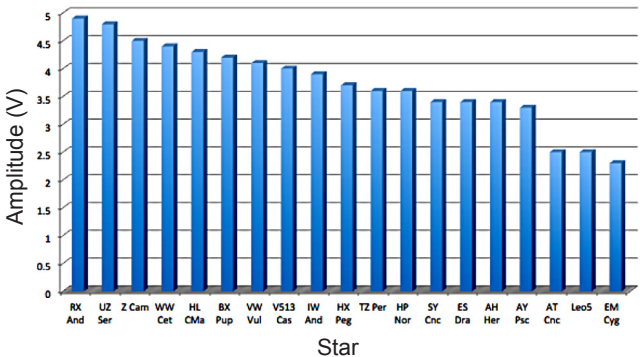

Figure 23. Outburst amplitudes (V magnitudes) for Z Cam stars.

4.5. VY Scl-like fading episodes

VY Sculptoris stars are CVs that behave much like NLs at maximum light; they may vary by up to one magnitude and they show no outbursts. Occasionally VY Scl stars undergo random fadings of two magnitudes or more. These episodes can last from days to years. Some Z Cam stars also exhibit dramatic fadings in their light curves, where they can bottom out at magnitudes fainter than their normal range (Figure 24).

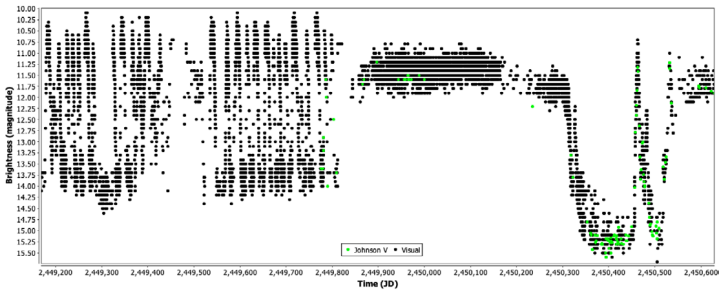

Figure 24. This AAVSO light curve of RX And shows two fading episodes. The short one (on the left) lasts about two months at about the normal magnitude range at minimum. The deeper fade (on the right) lasts 4.5 months and repeats after a brief outburst to maximum.

## 5. Misclassified Z Cams

The literature, CV catalogs, and the historical record are littered with stars classified as Z Cams or possible Z Cams, often based on slim, or no evidence. Some of these misclassified stars have been assumed to be Z Cams for a half-century or more. In fact, some have been touted as prototypical stars of the class and were used in studies characterizing the class of Z Cam stars. While many of these stars exhibit some of the characteristics, none of them are Z Cams.

As with the bona fide Z Cam stars, the entire sets of AAVSO data for each star were downloaded and the light curves were examined visually, both in gross, multi-year displays, and in close detail, cycle by cycle. Start dates of the data set and the beginnings and endings of outbursts could be determined by selecting individual data points on the light curve, displaying the detailed observational information in vstar.

Average cycle times were determined using a custom designed tool in vstar. Maxima could be hand-selected individually, and the average time between selections and mean magnitude were automatically calculated. The averages of the densest data sets were then calculated to derive the mean magnitude of maxima and the mean time between selections. The stars are listed below in order of R.A. Each star's position, maximum and minimum, and type are given in Table 2.

*TW Tri*  There are no indications of standstills in the AAVSO data from August 1995 to the present. This star is a UGSS with an average cycle time of 22 days and amplitude of ~3.5 magnitudes in V.

*KT Per*  Long considered a proto-typical Z Cam star, there are no indications of standstills in the AAVSO data from September 1967 to the present. This star is a UGSS with a ZZ Cet white dwarf. It has an average cycle time of 16.5 days and amplitude of ~4 magnitudes in V.

*AM Cas*  There are no indications of standstills in the AAVSO data from September 1988 to the present. This star is a UGSS with an average cycle time of 21.4 days and amplitude of ~4 magnitudes in V.

Table 2: Variable stars misclassified as Z Cam–type in the historical record.

| Name | R.A. (2000) h m s | Dec. (2000) ° ′ ″ | Max | Min | Type |
|---|---|---|---|---|---|
| TW Tri    | 01 36 37.01 | 32 00 39.9  | 13.3p  | 17.0p   | UGSS    |
| KT Per    | 01 37 08.78 | 50 57 20.3  | 10.6V  | 16.1V   | UGSS+ZZ |
| AM Cas    | 02 26 23.38 | 71 18 31.5  | 12.3p  | 15.2p   | UGSS    |
| FO Per    | 04 08 34.98 | 51 14 48.2  | 11.8V  | 16V     | UGSS    |
| AQ Eri    | 05 06 13.12 | –04 08 07.3 | 12.5p  | 16.5p   | UGSU    |
| BI Ori    | 05 23 51.77 | 01 00 30.6  | 13.2p  | 16.7p   | UGSS    |
| FS Aur    | 05 47 48.36 | 28 35 11.1  | 14.4p  | 16.2p   | UGSU:   |
| CN Ori    | 05 52 07.79 | –05 25 00.5 | 11.9v  | 16.3v   | UGSS    |
| V344 Ori  | 06 15 18.95 | 15 30 59.3  | 14.2p  | 17.5:p  | UG      |
| SV CMi    | 07 31 08.40 | 05 58 48.4  | 13.0p  | 16.9V   | UGSS    |
| UY Pup    | 07 46 31.25 | –12 57 09.1 | 13.5V  | 15.8V   | UG      |
| SW Crt    | 11 52 25.41 | –24 31 01.9 | 15.4CV | 16.4CV  | RRAB    |
| CG Mus    | 12 20 13.08 | –74 13 14.9 | 15.9p  | 17.0p   | RRAB    |
| V849 Her  | 16 35 45.72 | 11 24 58.1  | 15.0p  | 16.0p   | NL      |
| V391 Lyr  | 18 21 11.98 | 38 47 43.2  | 14.0p  | 17.0p   | UG      |
| V419 Lyr  | 19 10 13.91 | 29 06 14.0  | 14.4p  | <17.5p  | UGSU    |
| V1504 Cyg | 19 28 56.47 | 43 05 37.1  | 13.5p  | 17.4p   | UGSU    |
| FY Vul    | 19 41 29.95 | 21 45 59.0  | 13.4B  | 15.3B   | NL      |
| V1285 Cyg | 19 44 49.51 | 35 59 34.4  | 13.1p  | 14.8p   | SRB     |
| AB Dra    | 19 49 06.51 | 77 44 22.9  | 12.3V  | 15.8V   | UGSS    |
| V1363 Cyg | 20 06 11.53 | 33 42 37.6  | 13.0p  | <17.6p  | UGSU:   |
| EV Aqr    | 21 06 17.87 | 00 52 43.9  | 11.3v  | 13.6V   | SRA     |
| BS Cep    | 22 29 05.43 | 65 14 41.9  | 13.9p  | 16.0p   | UXOR    |
| AY Oct    | 23 27 51.00 | –75 40 40.7 | 15.0p  | 16.1p   | RRAB    |

*FO Per* There are no indications of standstills in the AAVSO data from October 1956 to the present. This star is a UGSS with an average cycle time of 10.3 days and amplitude of ~5 magnitudes in V.

*AQ Eri* This star is a known UGSU with an orbital period of 0.06094 day (Thorstensen *et al.* 1996).

*BI Ori* There are no indications of standstills in the AAVSO data from November 1978 to the present. This star is a UGSS with an average cycle time of 16 days and amplitude of ~4 magnitudes in V.

*FS Aur* This star is a highly unusual CV, possibly a triple system with a magnetic and freely precessing white dwarf (Tovmassian 2010). It has an orbital period of 0.0595 day. It is not a Z Cam.

*CN Ori* Considered a typical Z Cam star for decades, there are no indications of standstills in the AAVSO data from January 1931 to the present. This star is a UGSS with an average cycle time of 16 days and amplitude of ~4 magnitudes in V.

*V344 Ori*  There are no indications of standstills in the AAVSO data from October 1982 to the present. This star has a very long average cycle time of 443 days and an amplitude of ~5.5 magnitudes in V. This is either a UG or UGSU star.

*SV CMi*  Classified as a Z Cam in nearly every CV catalogue, there are no indications of standstills in the AAVSO data from December 1961 to the present. This star is a UGSS with an average cycle time of 20–24 days and amplitude of ~4.5 magnitudes in V.

*UY Pup*  There are no indications of standstills in the AAVSO data from February 1979 to the present. This star is a UGSS with an average cycle time of 57 to 60 days and amplitude of ~3.5 magnitudes in V.

*SW Crt*  This is an RRAB variable with a period of 0.493164 day (Figure 25).

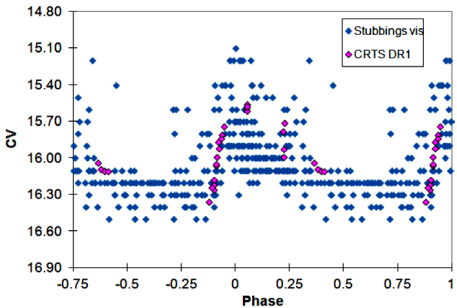

Figure 25. The light curve and period for SW Crt are from Otero (2012), derived from Rod Stubbings' visual data and the Catalina Real Time Survey data release 1 (Center for Advanced Computing Research 2013).

*V849 Her*  This is a nova-like variable, possibly of the VY Scl sub-type. It has an orbital period of 0.1414 to 0.0030 day (Ringwald *et al.* 2012).

*V391 Lyr*  There are no indications of standstills in the AAVSO data from April 1994 to the present. This star has a long average cycle time of ~110 days and an amplitude of ~4.2 magnitudes in V. V391 Lyr is a UG star.

*V419 Lyr*  This star is a known UGSU with an orbital period of 0.0864 day.

*V1504 Cyg*  This star is a known UGSU with an orbital period of 0.06951 day.

*FY Vul*  This star has an amplitude of only 1.5 magnitudes in V and quasi-periodic modulations with peaks on average every 16 to 24 days. This star is a NL, not a Z Cam.

*V1285 Cyg*  This is a red variable, spectral type M4IIIe, not a cataclysmic variable. It is a SRB varying irregularly between 13.1 and 14.8 p.

*AB Dra*  This star has been touted as a prototypical Z Cam since the 1960s, but there are no indications of standstills in the AAVSO data from August 1938 to the present. This star is a UGSS with an average cycle time of 10.5 days and amplitude of ~3.5 magnitudes in V.

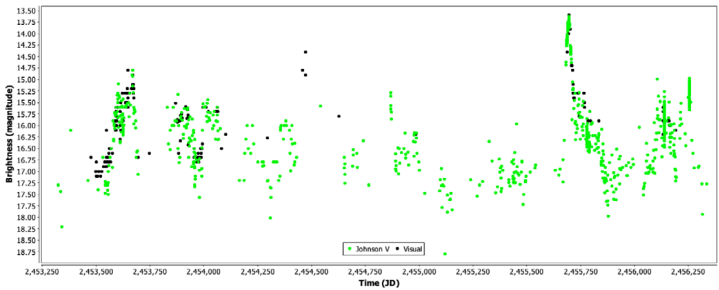

Figure 26. An AAVSO light curve of the long-term chaotic behavior of V1363 Cygni.

*V1363 Cyg*   This is an intriguing UG with an amplitude of 4.5 magnitudes in V that may turn out to be a UGSU, but it is not a Z Cam star (Figure 26).

*EV Aqr*   This is a red variable, a SRA varying between 11.2 and 13.8V with a period of 124.2 days.

*BS Cep*   This star is an UXOR, spectral type Ae, varying irregularly between 13.9 and 16.0p. It is not a UGZ.

*CG Mus*   Classified as a U Gem by Hoffmeister (1962) and as a possible UGZ in Downes and Shara (1993), this star was shown to be an RRAB with a period of 0.506815 day in Layden and Wachter (1997).

*AY Oct*   Previously listed as a possible UGZ in Downes and Shara (1993), this variable was mentioned as a possible RRAB in Cieslinski *et al.* (1998), but the identification of the star was in question. It is fairly certain the RR Lyrae star in the field is AY Oct and it has a period of 0.589918 day.

## 6. Concluding remarks

Of the sixty-five stars variously listed in the literature and catalogs as Z Cams or possible Z Cams, nineteen have been positively confirmed as members of the class. Twenty-four have been eliminated from the list, leaving twenty-two stars that require further investigation. Of the remaining twenty-two stars, fourteen are very likely to be NL or UGSS, but they cannot be ruled out with the existing data (see Table 3). Further long-term observations will be required. The remaining seven hold some promise. PY Per and ST Cha both have intriguing light curves that need to be enhanced before a valid conclusion can be made. V426 Oph has many UGZ-like features in its light curve, but is probably an intermediate polar (Hellier *et al.* 1990; Homer *et al.* 2004; Ramsay *et al.* 2008). The remaining four stars, V868 Cyg, V1404 Cyg, V1505 Cyg, and MN Lac, are quite faint at minimum and will require extra effort to monitor deeply enough and densely enough to make a determination of their true nature and classification. From the data currently at hand in the AAVSO International Database, V868 Cyg looks like it could be another unusual star, similar to IW And and V513 Cas.

Table 3. Remaining Suspected Z Cams.

| Name | R.A. (2000) h m s | Dec. (2000) ° ′ ″ | Max | Min | Preliminary Type |
|---|---|---|---|---|---|
| HS 0139+0559 | 01 41 39.93 | 06 14 37.5 | 15.2 | — | NL: |
| HS 0229+8016 | 02 35 58.20 | 80 29 44.2 | 13.7 | 15 | NL:/VY: |
| V392 Per | 04 43 21.39 | 47 21 25.8 | 15.2p | <17.5p | UG: |
| V368 Per | 02 47 32.70 | 34 58 27.4 | 15.2p | <17.5p | UG: |
| PY Per | 02 50 00.15 | 37 39 22.2 | 13.8p | 16.5p | UGZ: |
| HS 0642+5049 | 06 46 19.60 | 50 45 49.3 | 15.6 | — | NL |
| WZ CMa | 07 18 49.20 | –27 06 43.2 | 14.5p | 18.3p | UGSS: |
| ST Cha | 10 47 15.61 | –79 28 06.9 | 12.4V | 15.3V | UGZ: |
| V735 Sgr | 17 59 51.83 | –29 33 55.5 | 13.5p | 16.5p | ISB: |
| V426 Oph | 18 07 51.69 | 05 51 47.9 | 11.5V | 13.5V | UGZ:/DQ: |
| BP CrA | 18 36 50.89 | –37 25 53.6 | 13.5v | 15.9v | UGZ: |
| HS 1857+7127 | 18 57 20.36 | 71 31 18.8 | 13.9v | 17.2V | UGSS+E: |
| V868 Cyg | 19 29 04.50 | 28 54 26.0 | 14.3p | <17.8p | UGZ: |
| V1505 Cyg | 19 29 49.00 | 28 32 54.0 | 15.2p | <17.5p | UGZ: |
| V991 Aql | 19 35 34.84 | 06 33 45.8 | 14p | 16p | UXOR: |
| V1101 Aql | 20 13 04.07 | 15 35 46.8 | 14.3V | 14.7V | CV: |
| IS Del | 20 31 09.58 | 16 23 08.8 | 15.0p | <17.5p | UGSS: |
| TT Ind | 20 33 37.08 | –56 33 44.6 | 12.9V | 18.3V | UG: |
| HS 2133+0513 | 21 35 59.30 | 05 27 00.0 | 15.2V | <19.9V | NL/VY: |
| V1404 Cyg | 21 57 16.39 | 52 12 00.5 | 15.1V | 20.1:V | UGZ: |
| MN Lac | 22 23 04.63 | 52 40 58.9 | 15.1p | <18.0p | UGZ: |
| HS 2325+8205 | 23 26 50.29 | 82 22 11.2 | 13.8V | 17.8V | UG+E:/UGZ+E: |

This study has weeded out the imposters from the real Z Cams often cited in the literature and CV catalogues of the past. It may now be worthwhile to re-investigate conclusions drawn in earlier studies of Z Cam characteristics based on samples that contained stars that were not Z Cam stars (Meyer and Meyer-Hofmeister 1983; Szkody and Mattei 1984; Shafter *et al.* 2005). Today's catalogues, such as Ritter and Kolb (2003), should be revised to reflect our current understanding of these fascinating and complex systems.

It is significant that we have reduced the number of known Z Cams from the often stated "30 members or so" (Buat-Menard *et al.* 2001) to less than twenty. This is an extremely small percentage of the thousands of known CVs, and may indicate that Z Cam stars represent a brief stage in CV evolution.

In the context of CV evolution, the "hibernation scenario" (Shara *et al.* 1986) suggests that all dwarf novae will eventually become classical novae or have already experienced a classical nova eruption in the past. It has now been shown that Z Cam and AT Cnc, both bona fide members of the Z Cam class, have classical nova shells (Shara *et al.* 2007, 2012a, 2012b), demonstrating that at least some classical novae have evolved into dwarf novae.

It is clear we still lack an explanation of the root cause of the increased mass accretion rates that trigger standstills, and what ends them. Perhaps with a more homogeneous sample to model, the answers will be found. Having cleared away the imposters, we can now more effectively redouble our observational energies on the genuine class members, and gather the remaining data needed to do a meaningful study of this small population of CVs, including orbital periods, spectra in high, low, and standstill states, masses and radii of the primary and secondary components of these systems, and the systematic search for other classical novae shells around Z Cam stars.

**7. Acknowledgements**

The authors acknowledge with thanks the variable star observations from the AAVSO International Database contributed by observers worldwide and used in this research. Special thanks go to David Benn for his technical assistance and support. The authors also wish to thank Arne Henden, Steve Howell, and Paula Szkody for their guidance and encouragement, and the anonymous referee whose helpful suggestions made the paper more complete.